\documentclass[twocolumn]{svjour3} 

\smartqed

\usepackage{graphicx}
\usepackage{amsmath,amsfonts}

\begin{document}

\title{The spontaneous exchange bias effect on La$_{1.5}$Ba$_{0.5}$CoMnO$_{6}$}

\author{M. Boldrin        \and
        L. T. Coutrim     \and
        L. Bufai\c{c}al
}

\institute{Instituto de F\'{i}sica, Universidade Federal de Goi\'{a}s, 74001-970 , Goi\^{a}nia, GO, Brazil \\
              Tel.: \\
              Fax: \\
              \email{lbufaical@ufg.br}           %  \\
           \and
           L. Bufai\c{c}al - correspondence author 
}

\date{Received: date / Accepted: date}

\maketitle

\begin{abstract}

La$_{1.5}$Ba$_{0.5}$CoMnO$_{6}$ is a re-entrant cluster glass material exhibiting a robust negative exchange bias (EB) effect even after being cooled from an unmagnetized state down to low temperature in zero magnetic field. Here we thoroughly investigate this phenomena by performing magnetization as a function of applied field [$M(H)$] measurements at several different temperatures and maximum applied magnetic fields ($H_m$). The spontaneous EB (SEB) effect is observed below 20 K, and shows a maximum value for $H_m$ = 75 kOe. The effect is greatly enhanced when the $M(H)$ curves are measured after the system is cooled in the presence of a magnetic field. The asymmetry of the $M(H)$ curves here investigated can be well described by a recently proposed model based on the unconventional relaxation of the SG-like moments during the hysteresis cycle.

\end{abstract}

\section{Introduction}

There is a great interest in the exchange bias (EB) effect due to its potential applicability in high-density magnetic recording, giant magnetoresistance and spin valve devices. Conventionally, the unidirectional anisotropy (UA) across a ferromagnetic (FM)-antiferromagnetic (AFM) interface is set by cooling the system in the presence of an external magnetic field ($H$) \cite{Nogues}. Conversely, for the spontaneous EB (SEB) effect, the asymmetry in the hysteresis loop is observed at low temperatures ($T$) even after the system is cooled in zero $H$ \cite{Wang,Maity,Nayak,PRB2016}. For many of the materials presenting this recently discovered phenomena the presence of a re-entrant spin-glass (RSG) state, where a spin-glass (SG)-like phase is concomitant to other conventional magnetic phases \cite{Mydosh}, seem to play a key role \cite{Model,Model2}. In this context, the double-perovskite (DP) compounds are prospective candidates to exhibit SEB since its intrinsic structural disorder usually lead to competing magnetic interactions and frustration, the key ingredients for the emergence of SG-like behavior \cite{Mydosh}. 

From the SEB materials discovered so far, the great majority are perovskites \cite{Maity,PRB2016,Giri,Li,Murthy,JMMM2017}. The La$_{1.5}$Sr$_{0.5}$CoMnO$_{6}$ (LSCMO) DP stands out as presenting the largest SEB already known \cite{Murthy}, whilst for the sister compound La$_{1.5}$Ca$_{0.5}$CoMnO$_{6}$ (LCCMO) the effect is very subtle \cite{JMMM2017}. Although the presence of SG-like phase is stablished as requisite for the observation of the phenomena in LSCMO, LCCMO and other related compounds \cite{Model,Model2}, the microscopic mechanisms responsible for it are not completely known yet. It is not clear, for instance, why similar compounds as LSCMO and LCCMO exhibit such different SEB effects, and also why the phenomena is not observed for some other RSG DP compounds. In order to get insight on this questions, a detailed investigation of new SEB materials is necessary.

In this work we thoroughly investigate the magnetic properties of La$_{1.5}$Ba$_{0.5}$CoMnO$_{6}$ (LBCMO) DP by means of several magnetization as a function of $H$ [$M(H)$] measurements, performed at different temperatures and maximum applied $H$ ($H_m$). Our results show that La$_{1.5}$Ba$_{0.5}$CoMnO$_{6}$ presents a robust SEB effect below 20 K, although not as large as that observed for LSCMO. There is a great increase in the shift of the $M(H)$ curves when  the system is cooled in the presence of a magnetic field, with the conventional EB (CEB) effect getting larger as the cooling field ($H_{FC}$) increases up to $\sim$40 kOe, after which there is a tendency of saturation of the effect. We checked wether a recently proposed model, based on the pinning and relaxation of SG-like moments during the $M(H)$ cycling, can describe the EB effect in LBCMO, and the results show a very good adequacy between the theoretical and experimental data. The EB effect of LBCMO is compared to that of LSCMO and LCCMO and discussed in terms of the structural and magnetic particularities of each compound.

\section{Experimental details}

Polycrystalline LBCMO was synthesized by conventional solid-state reaction, as described elsewhere \cite{PRB2019}. It grows in rhombohedral $R\bar{3}c$ space group. The $M(H)$ measurements were performed in both zero field cooled (ZFC) and field cooled (FC) modes using a Quantum Design PPMS-VSM magnetometer, at an $H$ sweep rate of 30 Oe/s. Prior to the ZFC experiments it was given a particular care to eliminate any small trapped field in the magnet, with the sample being demagnetized with oscillating field at room temperature from one measurement to another.

\section{Results and Discussion}

Due to the presence of Co$^{2+}$/Co${3+}$ and Mn$^{3+}$/Mn$^{4+}$ mixed valence states in LBCMO, it shows two FM transitions at 186 and 155 K attributed respectively to the  Co$^{2+}$--O--Mn$^{4+}$ and Co$^{3+}$--O--Mn$^{3+}$ superexchange interactions, and an AFM transition at 45 K most probably related to the Co$^{3+}$--O--Mn$^{4+}$ coupling \cite{PRB2019}. The cationic disorder at Co/Mn sites and the competing magnetic phases lead to the emergence of a cluster spin glass (CG) phase at $\sim$ 72 K, making of this a RSG system \cite{Mydosh}.

\begin{figure}
\includegraphics[width=0.48 \textwidth]{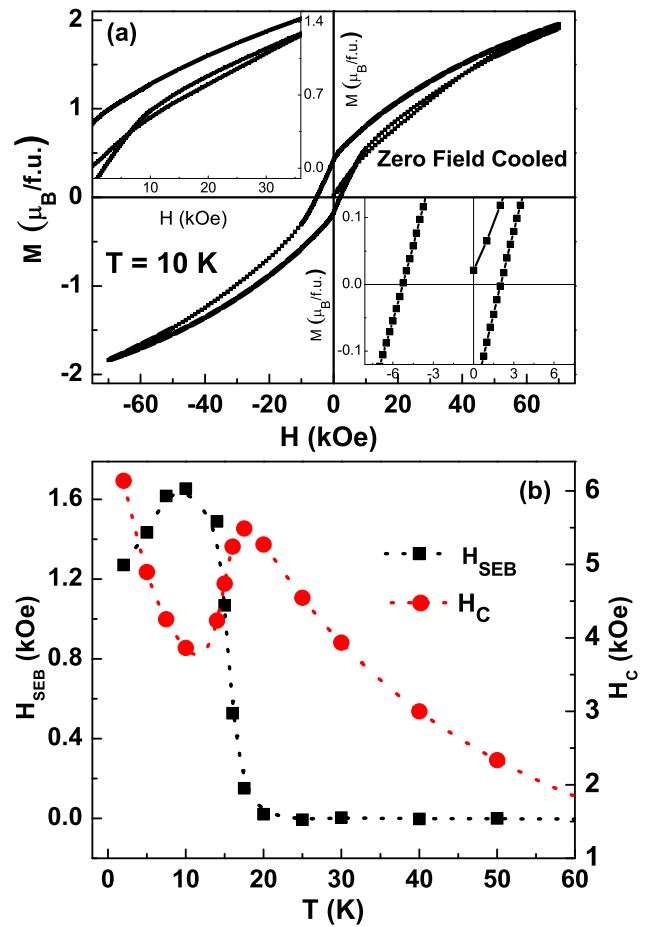}
\caption{(a) ZFC $M(H)$ curve of La$_{1.5}$Ba$_{0.5}$CoMnO$_{6}$ measured with $H_m$ = 70 kOe at 10 K. The upper inset shows a magnified view of the intersection between the virgin curve and the second branch of the loop, and the bottom inset highlights the loop shift toward left. (b) $H_{SEB}$ and $H_C$ evolution with $T$, where the doted lines are guides for the eye.}
\label{Fig_SEB}
\end{figure}

In this RSG system the anisotropic coercivity can be observed due to the combined action of FM and SG-like phases. The ZFC $M(H)$ measurements were performed for several $T$ at $H_m$= 70 kOe, and Fig. \ref{Fig_SEB}(a) shows a representative curve measured at 10 K. There is not a complete saturation of the magnetization at 70 kOe, due to the presence of a linear $H$-dependent AFM phase. The upper inset of Fig.\ref{Fig_SEB}(a) shows a magnified view of the first quadrant of the $M(H)$ loop, where it can be noticed that a stretch of the virgin curve lies outside the main loop. This is in general related to a $H$-induced irreversible magnetization process, and is a characteristic feature of several SEB materials, including the LSCMO and and LCCMO sister compounds \cite{Wang,Nayak,Murthy,JMMM2017}. 

The bottom inset of Fig. \ref{Fig_SEB}(a) highlights the shift of the $M(H)$ curve toward negative field direction, characterizing the SEB effect. The EB and the coercive fields are respectively defined as $H_{EB}=|H^{+}+H^{-}|/2$ and $H_{C}=(H^{+}-H^{-})/2$, where $H^{+}$ and $H^{-}$ represent respectively the right and the left cutoff fields. Fig. \ref{Fig_SEB}(b) displays the evolution of the ZFC EB field, $H_{SEB}$, as a function of $T$. It is interesting to note that $H_{SEB}$ initially increases with $T$, showing a maximum $H_{SEB}$ = 1.65 kOe at 10 K. This is different from the behaviour found for LSCMO and LCCMO sister compounds, for which $H_{SEB}$ monotonically decreases with increasing $T$ \cite{Murthy,JMMM2017}. A somewhat similar evolution of $H_{SEB}$ with $T$ is observed for La$_{1.5}$Sr$_{0.5}$Co$_{0.4}$Fe$_{0.6}$MnO$_{6}$, BiFeO$_{3}$-Bi$_{2}$Fe$_{4}$O$_{9}$ and Mn$_{3.5}$Co$_{0.5}$N \cite{Maity,Xie}, where the increased thermal energy enhances the alignment of the SG-like and AFM moments toward the field direction, enhancing its pinning with the FM phase and consequently increasing $H_{SEB}$. In the case of LBCMO, after the local maxima at $\sim$10 K some of the spins get energy enough to overcome the magnetic coupling as $T$ further increases and $H_{SEB}$ decreases up to 20 K, when it vanishes. Fig. \ref{Fig_SEB}(b) also displays the $H_C$ evolution with $T$. The resulting curve shows a peak at the $T$-region of descending $H_{SEB}$. This is a characteristic feature of EB systems, where the
gain in thermal energy favours some AFM spins to be ``dragged" by the rotating FM clusters during the $H$ cycling \cite{Nogues}, thus confirming that the effect here described is intrinsic.

\begin{figure}
\includegraphics[width=0.48 \textwidth]{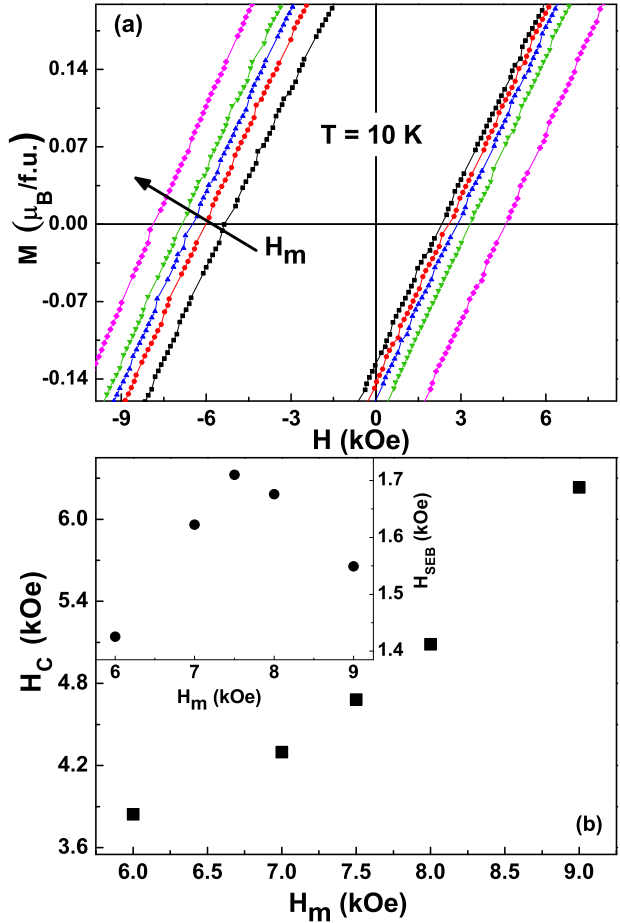}
\caption{(a) Magnified view of the cutoff fields of $M(H)$ loops carried with different $H_m$, at 10 K. The arrow indicates the direction of increasing $H_m$. (b) $H_C$ as a function $H_m$. The inset shows $H_{SEB}$ as a function of $H_m$.}
\label{Fig_Hm}
\end{figure}

We have also investigated the influence of $H_m$ on the SEB effect. Fig. \ref{Fig_Hm}(a) shows a magnified view of $M(H)$ curves measured at 10 K with different $H_m$, where it can be noticed that both the positive and negative coercive fields increase with increasing $H_m$. The SEB effect observed in DP compounds generally results from a delicate balance between the competing magnetic phases present in the system \cite{PRB2016,JMMM2017}. At one hand, the increase of $H_m$ may act to enhance the FM and/or SG-like phases which in turn will lead to the increase of $H^{-}$ and to the consequent increase of $H_C$ observed in Fig. \ref{Fig_Hm}(b). On the other hand, a large $H_m$ applied during the hysteresis cycle may drag the pinned spins toward the negative field direction when the system approaches $H_m$, resulting in the increase of $H^{+}$. As the inset of Fig. \ref{Fig_Hm}(b) shows, there is a $H_m$ value for which the difference between $H^{-}$ and $H^{+}$ is maximum, resulting in a maxima in the curve of $H_{SEB}$ as a function of $H_m$.

\begin{figure}
\includegraphics[width=0.48 \textwidth]{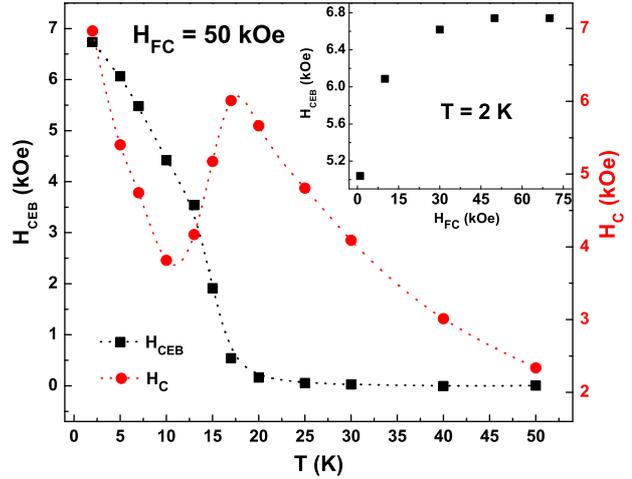}
\caption{$H_{CEB}$ and $H_C$ evolution with $T$, for $M(H)$ curves measured with $H_{FC}$ = 50 kOe. The inset shows $H_{CEB}$ as a function of $H_{FC}$ for $M(H)$ loops measured at 2 K, $H_{m}$ = 70 kOe.}
\label{Fig_CEB}
\end{figure}

There is a great increase in the EB effect when the system is cooled in the presence of an external field. The EB field obtained from $M(H)$ curves carried after the conventional process of FC the sample is herein called $H_{CEB}$. Fig. \ref{Fig_CEB} shows the evolution of $H_{CEB}$ and $H_C$ as a function of $T$ for $M(H)$ curves measured at $H_m$ = 70 kOe with a cooling field $H_{FC}$ = 50 kOe. The shape of the $H_{CEB}$ as a function of $T$ curve is remarkably different from that observed for $H_{SEB}$. This is because the external field favours the pinning of CG and AFM moments to the FM phase already at high $T$, resulting in a very large EB effect. With increasing $T$, the thermal energy gained get increasingly enough to overcome the magnetic coupling, resulting in the monotonically decrease of $H_{CEB}$ with increasing $T$.

As for other SEB materials discovered so far, the LBCMO compound presents RSG behavior at low $T$ \cite{PRB2019}. Thus, we have checked whether its UA could be described by a theoretical model recently proposed to explain the SEB effect in LCMO and LSCMO \cite{Model,Model2}. The model is based on the pinning of SG-like moments and on their unusual temporal evolution under the influence of the linear varying $H$ during a $M(H)$ cycle. In a $M(H)$ cycle the magnetization ($M$) depends on $H$, which in turn depends on time ($t$). Thus, the hysteresis curves can be displayed in the form of $M$ as a function of $t$ [$M(t)$]. Since the model is based on the temporal evolution of the magnetization of the glassy magnetic phase present in the system, the $M(t)$ form of presentation is most appropriate for checking the adequacy of the model to our results. 

\begin{figure}
\includegraphics[width=0.48 \textwidth]{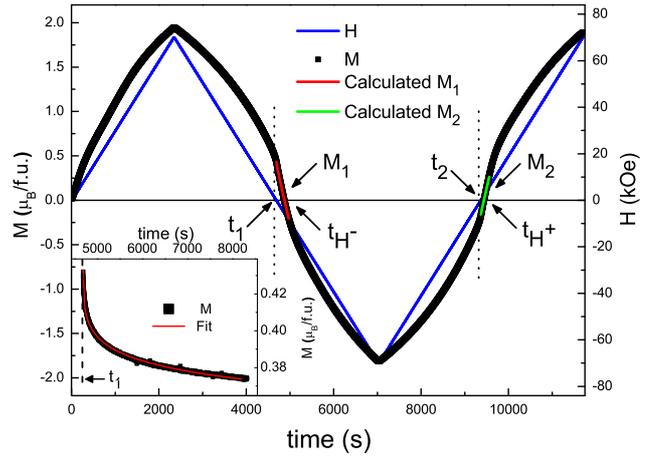}
\caption{$M(t)$ presentation of the hysteresis loop measured at 10 K, $H_m$ = 70 kOe. Red and green solid lines are the calculated $M_1$ (Eq. \ref{Eq1}) and $M_2$ (Eq. \ref{Eq2}) stretches, respectively. The blue solid line shows $H$ as a function of $t$. The inset shows the ZFC $TRM$ curve obtained at 10 K, after $H$ = 70 kOe was applied and subsequently turned off. The red line represent the best fit with the first brace of Eq. \ref{Eq1}.}
\label{Fig_model}
\end{figure}

Fig. \ref{Fig_model} presents the same $T$ = 10 K and $H_m$ = 70 kOe $M(H)$ cycle shown in fig. \ref{Fig_SEB}(a), but now displayed in the $M(t)$ form. $t_1$ and $t_2$ represent the times when $H$ = 0, while $t_{H^-}$ and $t_{H^+}$ correspond to the instants when $M$ = 0. The main goal of the proposed phenomenological model is to describe the $H_{EB}$ effect by calculating the $M_1$ and $M_2$ stretches of the $M(H)$ curve that encompass the $H^{-}$ and $H^{+}$ coercive fields, respectively. For the $M_1$ stretch, it is considered that in the time-interval $t_1 \leq t \leq t_{H^{-}}$ the SG-like moments are relaxing due to the positive $H_{m}$ firstly applied, although the system is already under the effect of a linearly varying negative $H$. Therefore, the equation describing the $M_1$ stretch is
\begin{equation}
\begin{split}
M_{1}(t)& = \{M_{sp} + M_{0}e^{-\left[(t-t_{1})/t_{p}\right]^{n}}\} \\
& - \{A(t-t_{1}) + B(t-t_{1})^{r}\}, \label{Eq1} \\
\end{split}
\end{equation}
where the first pair of braces represents the magnetic relaxation of the SG-like phase due to the  previously applied positive $H_m$, and the second pair of braces account for the contributions of the AFM and FM phases to $M_1$ when under the effect of the immediately applied negative $H$. The $A$ parameter is related to the linear dependence of the AFM phase with $H$ (and consequently with $t$), while the $B$ and $r$ parameters account for the non-linear contribution of the FM phase to the magnetization \cite{Model,Model2}. 

Prior to applying Eq. \ref{Eq1} to describe the $M_1$ stretch, the temporal evolution of the SG-like phase must be computed. This is done by calculating the parameters of the first brace of Eq. \ref{Eq1} separately. It must be noticed that the terms in the first pair of braces correspond to the stretched exponential equation that describes the thermo-remanent magnetization ($TRM$) of RSG systems \cite{Mydosh,Chamberlin}, where $M_{sp}$ represents the spontaneous magnetization of the conventional phase (FM in this case), $M_0$ is the initial magnetization of the SG-like phase at the instant $t_1$ when $H$ = 0 [see Fig. \ref{Fig_model}], and $t_p$ and $n$ are the time and the time-stretch exponential, respectively. The fitting of the ZFC $TRM(t)$ curve obtained for LBCMO at 10 K, after an $H_m$ = 70 kOe being applied and subsequently turned off, is displayed in the inset of Fig. \ref{Fig_model}. The best fit yields $M_{sp}$ = 0.218 $\mu_{B}$/f.u., $M_0$ = 0.235 $\mu_{B}$/f.u., $t_p$ = 8.15$\times$10$^{5}$ s and $n$ = 0.164, these values lying in between those observed for LSCMO and LCCMO \cite{Model}.

Now the $M_1$ stretch can be fitted by Eq. \ref{Eq1}, where the parameters of the first brace are kept fixed at the values obtained from the fit of the $TRM$ curve. The red solid line in Fig. \ref{Fig_model} shows the best calculated curve, resulting in $A$ = 4.87$\times$10$^{-4}$ $\mu_{B}$/f.u., $B$ = 1.96$\times$10$^{-3}$ $\mu_{B}$/f.u. and $r$ = 0.97. 

For the $M_2$ stretch, it is assumed that a part of the SG-like moments are relaxing due to the negative field previously applied while the others are still relaxing due to the firstly applied positive $H_m$, \textit{i.e.}, they are still pinned toward the positive direction. This results in the following equation
\begin{equation}
\begin{split}
M_{2}(t)& = -\{M_{sp} + xM_{0}e^{-\left[(t-t_{2})/t_{p}\right]^{n}}\} + \\
& \{(1-x)M_{0}e^{-\left[(t-t_{1})/t_{p}\right]^{n}}\} + \{A(t-t_{2}) + B(t-t_{2})^{r}\}, \label{Eq2} \\
\end{split}
\end{equation}
where the first pair of braces represents the decay of the SG-like spins that are relaxing from the negative field applied before, the second pair corresponds to the relaxation from the positive field firstly applied, and the third pair represents the variation in the AFM/FM phases due to the just applied positive field. All the parameters of Eq. \ref{Eq2} are kept fixed with the values obtained from the fit of $M_1$, with the exception of the $x$ parameter that measures the amount of SG-spins that has been flipped toward negative direction due to the $H$ cycling. 

The resulting curve obtained from Eq. \ref{Eq2} is displayed as the green solid line in Fig. \ref{Fig_model}, which tells that $\sim$74$\%$ of the SG-like moments are still pinned toward positive $H$ direction at the $M_2$ patch, resulting in a small $H^{+}$ and consequently in a large $H_{SEB}$. In the case of $M(H)$ measurements performed after the FC process, it is expected an even smaller portion of SG-like spins flipping toward the negative direction during the cycle, since in this case a positive field favours the pinning of the spins already from above $T_C$, resulting in the very large CEB effect observed.

\section{Conclusions}

In summary, in this work we thoroughly investigated the EB effect in a polycrystalline sample of LBCMO DP. This is a RSG material for which the glassy magnetic behaviour is manifested at low $T$ as a consequence of competing Co$^{2+}$--Mn$^{4+}$ and Co$^{3+}$--Mn$^{3+}$ FM phases and Co$^{3+}$--Mn$^{4+}$ AFM phase. The $H_{SEB}$ = 1.27 kOe value observed at 2 K, measured with $H_m$ = 70 kOe, is intermediate between the 3.3 kOe and the 0.25 kOe values reported for LSCMO and LCCMO, respectively. This can be understood in terms of the crystal structure of each compound and the details of the magnetic relaxation of each sample under the influence of the varying field at the $M(H)$ cycle. When the system is cooled in the presence of an external field, the EB effect is greatly enhanced. The UA set spontaneously in LBCMO after a ZFC protocol could be well described by a model recently proposed to explain such effect in terms of the pinning and relaxation of SG-like moments during the $M(H)$ cycle.

\begin{acknowledgements}
This work was supported by Conselho Nacional de Desenvolvimento Cient\'{i}fico e Tecnol\'{o}gico (CNPq) [No. 400134/2016-0], Funda\c{c}\~{a}o de Amparo \`{a} Pesquisa do Estado de Goi\'{a}s (FAPEG) and Coordena\c{c}\~{a}o de Aperfei\c{c}oamento de Pessoal de N\'{i}vel Superior (CAPES).
\end{acknowledgements}

\end{document}